\documentclass[reprint]{iopart}

\usepackage[utf8]{inputenc}

\usepackage{iopams}
\usepackage{setstack}

\usepackage{graphicx}

\graphicspath{{Figures/}{/}}
\usepackage{color}
\definecolor{darkblue}{rgb}{0., 0., 0.55}
\usepackage[colorlinks=true, urlcolor=darkblue, citecolor=darkblue, linkcolor=darkblue]{hyperref}

\usepackage{dsfont}

\usepackage{xcolor}
\usepackage{ulem}

\newcommand{\rbold}{\ensuremath{\boldsymbol{r}}}
\newcommand{\ud}{\ensuremath{\mathrm{d}}}
\newcommand{\diag}{\ensuremath{\mathrm{diag}}}
\newcommand{\ue}{\ensuremath{\mathrm{e}}}
\newcommand{\ui}{\ensuremath{\mathrm{i}}}

\newcommand{\bmap}[1][\ensuremath{n}]{\ensuremath{B_{#1}}}
\newcommand{\qbmap}[1][n]{\ensuremath{\mathcal{B}_{#1}}}
\newcommand{\qbmapr}[1][n,\rbold]{\ensuremath{\mathcal{B}_{#1}}}
\newcommand{\qbmaprrandom}[1][n,\rbold,\levelp,\levelq]{\ensuremath{\mathcal{B}_{#1}}}
\newcommand{\qmaprrandom}[1][\rbold]{\ensuremath{\mathcal{U}_{#1}}}
\newcommand{\qrefl}[1][\rbold]{\ensuremath{\mathcal{R}_{#1}}}
\newcommand{\op}[1][N]{\ensuremath{\mathrm{op}_{#1}}}
\newcommand{\torus}{\ensuremath{\mathbb{T}}}
\newcommand{\fourier}{\ensuremath{\mathcal{F}}}

\newcommand{\mrandom}[1][\mathrm{cue}]{\ensuremath{\mathcal{M}_{#1}}}
\newcommand{\urandom}{\ensuremath{\mathcal{U}^{\mathrm{cue}}_{\levelp, \levelq} }}
\newcommand{\urandomoo}{\ensuremath{\mathcal{U}^{\mathrm{cue}}_{\levelp=0, \levelq=0} }}
\newcommand{\ufourier}[1][\levelq]{\ensuremath{\mathcal{U}_{#1}}}
\newcommand{\hus}[1][\psi]{\ensuremath{\mathcal{H}_{#1}}}

\newcommand{\ncue}{\ensuremath{N_{\mathrm{cue}}}}

\newcommand{\nsub}{\ensuremath{N_{\mathrm{sub}}}}
\newcommand{\nrect}{\ensuremath{N_{\mathrm{rect}}}}

\newcommand{\bshift}[1][\ensuremath{n}]{\ensuremath{S_{#1}}}
\renewcommand{\bshift}[1][]{\ensuremath{S}{#1}}

\newcommand{\gnat}{\ensuremath{\gamma_{\mathrm{nat}}}}
\newcommand{\ginv}{\ensuremath{\gamma_{\mathrm{inv}}}}
\newcommand{\gtyp}{\ensuremath{\gamma_{\mathrm{typ}}}}
\newcommand{\refl}{\ensuremath{r}}
\newcommand{\intens}{\ensuremath{\mathbb{R}_{+}}}

\newcommand{\muL}{\ensuremath{\mu_\mathrm{L}}}
\newcommand{\mulevel}{\ensuremath{\mu_{\gamma}^{\levelp,\levelq}}}
\newcommand{\muqm}{\ensuremath{\mu_{\mathrm{qm}}}}
\newcommand{\mucl}{\ensuremath{\mu_{\mathrm{cl}}}}

\newcommand{\jsd}{\ensuremath{d_{\mathrm{JS}}}}
\newcommand{\entropy}{\ensuremath{H}}

\newcommand{\level}{\ensuremath{\ell}}
\newcommand{\levelp}{\ensuremath{\level_p}}
\newcommand{\levelq}{\ensuremath{\level_q}}

\newcommand{\recta}{\ensuremath{\alpha}}
\newcommand{\brecta}{\ensuremath{\boldsymbol{\alpha}}}

\newcommand{\eqref}[1]{(\ref{#1})}
\newcommand{\EQref}[1]{Eq.~\eqref{#1}}

\begin{document}

\title[Local random vector model for chaotic resonance states]
   {Local random vector model
    for semiclassical fractal structure of chaotic resonance states}

\author{Konstantin Clau\ss$^{1,2}$, Roland Ketzmerick$^{1}$}
\address{%
    $^1$Technische Universit\"at Dresden, Institut f\"ur Theoretische Physik and Center for Dynamics, 01062 Dresden, Germany
}\vspace{5pt}%

\address{$^2$Department of Mathematics, Technical University of Munich, Boltzmannstr. 3, 85748 Garching b.~M\"unchen, Germany}

\vspace{10pt}
\begin{indented}
\item \today \date{\today}
\end{indented}

\begin{abstract}
The semiclassical structure of resonance states of classically chaotic 
scattering systems with partial escape is investigated.
We introduce a local randomization on phase space for the baker map with escape,
which separates the smallest multifractal scale 
from the scale of the Planck cell.
This allows for deriving a semiclassical description of
resonance states based on a local random vector model and
conditional invariance.
We numerically demonstrate that the resulting classical measures
perfectly describe resonance states of all decay rates $\gamma$
for the randomized baker map.
By decreasing the scale of randomization
these results are compared to the deterministic baker map with partial escape.
This gives the best available description of its resonance states.
Quantitative differences indicate that a semiclassical
description for deterministic chaotic systems  must take into account that
the multifractal structures persist down to the Planck scale.
\end{abstract}
\submitto{\jpa}

\section{Introduction}
Understanding the correspondence of quantum and classical dynamics
is an essential task in complex systems \cite{HaaGnuKus2018}.
Most prominently, the semiclassical structure of quantum eigenstates in closed systems is related
to classically invariant phase-space structures
\cite{Ber1977b,Per1977a,Vor1979}.
For classically ergodic dynamics 
almost all quantum eigenstates converge weakly towards the uniform measure
on phase space as proven by the quantum ergodicity theorem
\cite{Shn1974,CdV1985,Zel1987,ZelZwo1996,BaeSchSti1998}.
This uniform limit is also established
for quantum maps \cite{DegGraIso1995,BouDeB1996,NonVor1998}.
The amplitude distribution of eigenstates is described
by random wave models, first introduced for quantum billiards \cite{Ber1977b}, in which a complex Gaussian distribution of wave
amplitudes leads to a universal
exponential distribution of the intensities
\cite{NonVor1998,McDKau1988,AurSte1991,LiRob1994,Pro1997b,%
      Izr1987,KusMosHaa1988,Bae2003}.
Extensions include constraints by two-point correlations
\cite{BieLepHel2003,UrbRic2003, UrbRic2006}
and systems with a mixed phase space \cite{BaeSch2002a,BaeNon:p}.

In contrast, 
for systems with loss of particles or intensity
in some interaction region \cite{AltPorTel2013}
the semiclassical limit is not fully understood.
Such scattering systems,
e.g., the three-disk system
\cite{GasRic1989c,Wir1999,WeiBarKuhPolSch2014} 
or optical microcavities \cite{CaoWie2015},
are described by resonance poles and the
corresponding resonance states $\psi$ which have a decay rate $\gamma$.
The distribution of decay rates is well described 
by a fractal Weyl law in systems with full escape
\cite{Sjo1990,Lin2002,
    LuSriZwo2003,SchTwo2004,RamPraBorFar2009,
    EbeMaiWun2010,ErmShe2010,
    PedWisCarNov2012,NonSjoZwo2014},
and has been studied in systems with partial escape
\cite{WieMai2008,NonSch2008,GutOsi2015,SchAlt2015}.
Here we focus on
the semiclassical structure of resonance states
in chaotic systems with escape.
They are described by multifractal measures on phase space
\cite{CasMasShe1999b,
    LeeRimRyuKwoChoKim2004,
    KeaNovPraSie2006,
    NonRub2007,
    KeaNonNovSie2008,
    NovPedWisCarKea2009,
    HarShi2015,
    KoeBaeKet2015,
    CarBenBor2016,
    KulWie2016,
    ClaKoeBaeKet2018,
    ClaAltBaeKet2019,
    BitKimZenWanCao2020}
with exponentially distributed intensity fluctuations \cite{ClaKunBaeKet2021}.
These measures are conditionally invariant under the classical dynamics with escape
\cite{DemYou2006,AltPorTel2013}
and strongly depend on the corresponding quantum decay rate $\gamma$ 
\cite{NonRub2007,ClaKoeBaeKet2018,ClaAltBaeKet2019}.
Such questions are also of interest in
mixed systems and in non-Hermitian PT-symmetric systems
\cite{KopSch2010,KoeMicBaeKet2013,MalPolSch2015,MudGra2020}.
For fully chaotic systems 
there exist approximative descriptions of resonance states
with arbitrary decay rates $\gamma$
for full \cite{ClaKoeBaeKet2018} and partial escape \cite{ClaAltBaeKet2019},
however, the precise semiclassical limit is not known.

The main difficulty in describing chaotic resonance states,
according to our understanding,
is the fact that their multifractal structures
persist all the way down to the scale of a Planck cell.
Since the multifractal structure has its origin in the classical dynamics,
there is no separation of the smallest classical scale
and the scale of a Planck cell,
not even when going to the semiclassical limit. 

In this paper we introduce a chaotic model system with escape
for which the smallest multifractal scale and the
scale of the Planck cell are separated.
This allows for deriving a semiclassical description of
resonance states.
Specifically, we define the randomized baker map with escape,
where in addition to the quantum baker map
and escape from phase space,
 local randomization operations on phase space
with an adjustable scale are applied.
Resonance states show multifractal structures on scales larger than the scale of
randomization and are uniformly fluctuating on smaller scales.
For partial escape, we derive classical measures, which
(i) are conditionally invariant 
with some desired decay rate $\gamma$
on scales larger than the scale of randomization
and (ii) obey an extremum principle derived from a
local random vector model in each randomization region.
We numerically demonstrate that for all decay rates $\gamma$
these classical measures 
perfectly describe resonance states
in the semiclassical limit.
By decreasing the scale of randomization
these results are compared to the deterministic baker map with partial escape.
This gives the best available description of its resonance states.
Quantitative differences indicate that a semiclassical
description for deterministic chaotic systems must take into account that
the multifractal structures persist down to the Planck scale.

The paper is organized as follows.
In Sec.~\ref{sec:classicalbaker}
we review the classical baker map and its symbolic dynamics.
Section~\ref{sec:quantumbaker}
introduces the randomized quantum baker map with escape and 
discusses the structure of its resonance states.
In Sec.~\ref{sec:semiclassics} we determine classical measures
from a local random vector model with the constraint of
conditional invariance. We conjecture
that these classical measures describe
the semiclassical structure of resonance states.
Numerical support for the conjecture is presented in
Sec.~\ref{sec:numerical}.
A comparison to the deterministic quantum baker map is made in
Sec.~\ref{sec:usualbaker}
showing qualitative similarity, but quantitative difference.
Section~\ref{sec:outlook} gives an outlook and in 
\ref{sec:rmt}
a random matrix model with escape is discussed.

This paper touches many topics that Fritz Haake has worked on
\cite{HaaGnuKus2018},
as it studies the semiclassical limit of a quantum model combining
classical deterministic chaotic dynamics with random matrices.
Specifically,
the local randomization procedure is related to the
blurring of phase-space resolution 
studied by Fritz \cite{WebHaaSeb2000,WebHaaBraManSeb2001,ManWebHaa2001}.
Furthermore the present paper deals with open quantum systems,
another field to which Fritz contributed substantially
in terms of scattering systems \cite{AlbHaaKurKusSeb1996,HacVivHaa2003},
as well as in terms of coupling to a thermal environment
\cite{Haa1973, StrHaa2003}.
It would have been a pleasure to discuss the results with Fritz
and to carefully listen to his insights.

\section{Classical baker map with escape}
\label{sec:classicalbaker}

In this section we introduce the baker map, one of
the paradigmatic models for chaos,
combined with escape from
phase space~\cite{LaiTel2011,AltPorTel2013}.
Its symbolic dynamics will be used to define a phase-space partition with rectangles that will
be relevant for the randomized quantum baker map with escape introduced in the next section.

\subsection{Baker map with escape}
\label{sec:classical-baker-escape}

The baker map with $n$ stripes is defined on the two-torus
$\torus = [0, 1) \times [0, 1)$ as follows \cite{ArnAve1968}.
Consider the vertical rectangles $A_k = [k/n, (k+1)/n) \times [0, 1)$ 
for $k \in \{0, \dots, n-1\}$. The $n$-baker map compresses
each of these rectangles along the $p$-direction
by the factor $1/n$ and stretches it
along the horizontal $q$-direction by the factor $n$, after which
they are stacked on top of each other,
\begin{eqnarray}
    \bmap : \torus \rightarrow \torus\\
     A_k \ni (q, p) \mapsto
     \big(q n - k, (p + k)/n\big).\nonumber
\end{eqnarray}
Note that $\bmap$ is volume preserving on $\torus$.

\begin{figure}[b!]
    \centering    \includegraphics[scale=1]{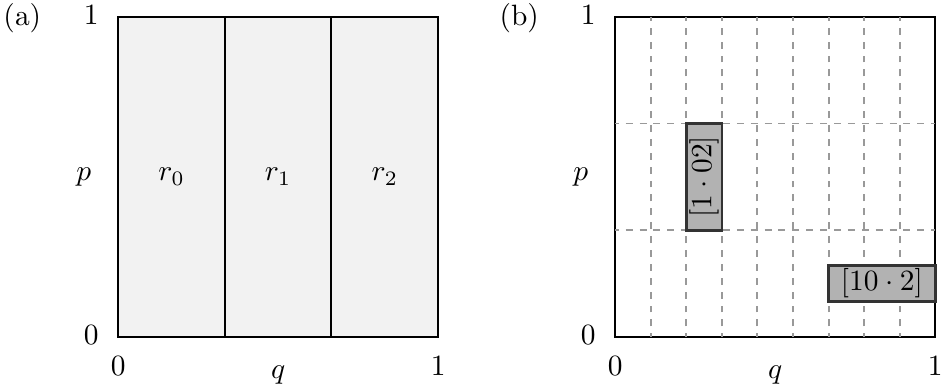}
    \caption{(a) Illustration of reflectivities  $\refl_k$ in rectangles $A_k$ 
        for the $3$-baker map with escape.
        (b) Phase-space partition by rectangles
        $[\boldsymbol{\recta}_{m_p,m_q}]$ with $m_p=1$ and $m_q=2$
        shown by dashed lines.
        The rectangle $[1 \cdot 02]$ and its iterate $[10 \cdot 2]$ are highlighted, where
        the latter intersects with $3$ rectangles of the partition.
    }
    \label{fig:sketch}
\end{figure}
For the baker map with escape we consider a loss of intensity which
is constant in each of the rectangles $A_k$. In particular,
let $\rbold\in \mathbb{R}^n_{+}$ and define the reflectivity function
$\refl : \torus \rightarrow \intens$ as
$A_k\ni (q, p) \mapsto \refl_k$, see Fig.~\ref{fig:sketch}(a).
Extending the phase space with the intensity space $\intens$,
the map with escape is defined by
\begin{eqnarray}
    \label{eq:map-with-escape}
\bmap[n,\rbold] : \torus \times \intens
\rightarrow \torus \times \intens \\
\big((q, p), I\big)\mapsto \big(\bmap(q, p), I\cdot\refl(q, p)\big),
\nonumber
\end{eqnarray}
i.e., the intensity in $A_k$ is changed by the factor $\refl_k$.
One distinguishes systems with full escape, where at least one of the $\refl_k = 0$,
from systems with partial escape, where all $\refl_k > 0$ \cite{AltPorTel2013}.
Often $\refl_k \leq 1$ is considered, leading to classical decay
under the forward map \cite{AltPorTel2013}.
On the other hand, considering $\refl_k \geq 1$ for some (or all)
$A_k$ implies growth of intensities \cite{AltPorTel2015}.
The simple structure 
of the baker map with escape allows for determining
the natural decay rates under forward and backward dynamics analytically.
The natural measure is uniform along the unstable $q$-direction
such that the natural decay rate is given by
$\gnat = -\ln ( \frac{1}{n} \sum_{k=0}^{n-1} \refl_k)$ \cite{AltPorTel2013b}. 
The natural measure of the backward dynamics is uniform along the stable
$p$-direction, such that the
natural growth rate of the inverse map is
$\ginv = \ln (\frac{1}{n} \sum_{k=0}^{n-1}
 \frac{1}{\refl_k})$ \cite{AltPorTel2015,ClaAltBaeKet2019}.
Another important classical decay rate is the average decay of
a typical ergodic orbit \cite{NonSch2008},
the so-called typical decay rate
$\gtyp = - \frac{1}{n}\sum_{k=0}^{n-1}  \ln \refl_k$.
One has the ordering $\gnat \leq \gtyp \leq \ginv$,
with equality if all $r_k$ are identical.

\subsection{Symbolic dynamics and phase-space partition}
\label{sec:symbolic-map}
The baker map has a complete symbolic dynamics in terms of the
Markov partition $\{A_k\}_{k=0}^{n-1}$.
Therefore it is equivalent to a shift on symbolic sequences, which
is recalled in the following.
Let $\bshift$ be the right shift operator on the
symbolic space $Z_n = \{0,1,\dots,n-1\}^\mathbb{Z}$, which 
consists of bi-infinite sequences
$\boldsymbol{\recta} = \{\alpha_i\}_{i=-\infty}^{\infty} = 
\dots \recta_{-2}\recta_{-1}\cdot
\recta_0\recta_1 \recta_2 \dots$, where $\alpha_i \in \{0, \dots, n-1\}$ for all $i \in \mathbb{Z}$.
The shift operator $\bshift : Z_n \rightarrow Z_n$ acts on such
sequences as
\begin{equation}
\bshift(\boldsymbol{\recta})
= \dots \recta_{-2}\recta_{-1}\recta_0\cdot\recta_1 \recta_2 \dots,
\end{equation}
i.e., it shifts the center of this sequence by one such that $(\bshift(\boldsymbol{\recta}))_i = \recta_{i + 1}$.
There is a mapping from the space of symbolic 
sequences to the torus, $J_n: Z_n \rightarrow \torus$,
\begin{equation}
    J_n(\boldsymbol{\recta}) = (q, p)
    := \left(\sum_{i=0}^\infty \frac{\recta_i}{n^{i + 1}} , \quad\sum_{i=0}^{\infty} \frac{\recta_{-(i + 1)}}{n^{i + 1}} \right),
\end{equation}
which is bijective on a subset of $Z_n$ with full (Lebesgue) measure
\cite{NonRub2007}.
Thus, $J_n$ conjugates the symbolic shift with the baker map, $\bmap = J_n \circ \bshift \circ J_n^{-1}$,
and they have equivalent dynamics.

The symbolic space allows to define a partition of the torus in terms of rectangles on phase space by finite sequences
$\boldsymbol{\recta}_{m_p,m_q} = \{\alpha_i\}_{i=-m_p}^{m_q - 1}$ of 
length $m_p + m_q$, see illustration in Fig.~\ref{fig:sketch}(b).
For such a finite sequence a subset
of $Z_n$ is defined as
\begin{equation}
    [\boldsymbol{\recta}_{m_p,m_q}]  :=
\{\boldsymbol{\beta}\in Z_n : \beta_i = \alpha_i \;\;\; \forall -m_p \leq i \leq m_q - 1\},
\label{eq:def-rectangle}
\end{equation}
which contains all infinite sequences with the central part
being equal to $\boldsymbol{\recta}_{m_p,m_q}$.
By construction  $J_n([\boldsymbol{\recta}_{m_p,m_q}])$ is a
rectangle on $\torus$ and a subset of $A_{\recta_0}$.
There are $n^{m_p + m_q}$ such non-overlapping rectangles of size
$\Delta p = 1/n^{m_p}$ and $\Delta q = 1/n^{m_q}$,
see Fig.~\ref{fig:sketch}(b).
In the following we will use the term rectangle for both, the sets of symbolic sequences
$[\boldsymbol{\recta}_{m_p,m_q}]$ and their  
image 
$J_n([\boldsymbol{\recta}_{m_p,m_q}])$
on phase space.

\section{Randomized quantum baker map with escape}
\label{sec:quantumbaker}

In this section we first give the 
standard quantization of the baker map with escape. 
Then we introduce a randomization operator that acts locally
on all phase-space rectangles, described by finite symbolic sequences,
as introduced in Sec.~\ref{sec:symbolic-map}.
The resonance states of this randomized baker map are multifractal down to the scale of randomization
and uniformly fluctuating on smaller scales.
Thus, the proposed randomization
separates the smallest 
classical scale from the scale of a Planck cell.
This allows for a well-defined semiclassical limit,
where the size of a Planck cell becomes arbitrarily small
compared to all classical scales.

\subsection{Quantum baker map with escape}

The baker map can be quantized with geometric quantization
\cite{BalVor1989,Sar1990}.
For this  the Hilbert space $\mathbb{C}^N$
with dimension $N \in n\mathbb{N}$ is considered and 
the time-evolution operator on  $\mathbb{C}^N$ is defined
in position basis as
\begin{equation}
    \qbmap := \fourier^{-1}_{N}\ 
    \diag (\fourier_{N/n} , \dots , \fourier_{N/n}).
\end{equation}
Here, the matrices $\fourier_M$ denote the discrete Fourier transform,
which for arbitrary $M$ are given by
\begin{equation}
    {(\fourier_M)}_{ml} := M^{-\frac{1}{2}}  \ue^{-\frac{2\pi \ui}{M} ( m+\frac{1}{2})(l+\frac{1}{2})}.
    \label{eq:def-fourier}
\end{equation}
Note that the exponents in the definition of $\fourier$ ensure
for the closed baker map that symmetries of $\bmap$ are respected
by the eigenfunctions of $\qbmap$ \cite{Sar1990}.

The quantization of the baker map with escape is composed of the closed quantum map $\qbmap$ and a reflectivity operator
$\qrefl$. For constant escape from the rectangles
$A_k$, as in the classical map, one has
\begin{equation}
    \qrefl := \diag(\sqrt{\refl_0}\,\mathds{1}_{N/n}, \dots, \sqrt{\refl_{n-1}} \, \mathds{1}_{N/n}).
    \label{eq:refl-operator}
\end{equation}
The quantization of the baker map with escape, Eq.~\eqref{eq:map-with-escape},
is then given by
\begin{equation}
    \qbmapr := \qbmap \qrefl,
    \label{eq:def-qbmap}
\end{equation}
which has been extensively studied for full and partial escape
\cite{KeaNovPraSie2006,NonRub2007,KeaNonNovSie2008,%
    NovPedWisCarKea2009,PedWisCarNov2012,CarBenBor2016}.

\subsection{Randomization}

We introduce a procedure to randomize the quantum dynamics
of the $n$-baker map.
As randomization regions we consider the rectangles
$[\boldsymbol{\recta}_{\levelp,\levelq}]$
from a finite symbolic sequence of length $\levelp$ in backward and $\levelq$
in forward direction, see Eq.~\eqref{eq:def-rectangle}.
The randomization within each of these $n^{\levelp + \levelq}$
rectangles is quantum mechanically achieved 
with random matrices $\mrandom[\ncue]$ 
of dimension $\ncue$,
where we choose the circular unitary ensemble (CUE) which has no symmetries.
Then the dimension $N$ of the full Hilbert space
is given by $N = n^{\levelq + \levelp} \ncue$.
The randomization operator in a basis corresponding to these rectangles is given by the block-diagonal operator
\begin{equation}
    \hat{\mathcal{U}}^{\mathrm{cue}}_{\levelp,\levelq} =
    \diag(\mrandom[\ncue], \dots, \mrandom[\ncue]),
\end{equation}
consisting of $n^{\levelq + \levelp}$ CUE-matrices with
dimension $\ncue$.
In order to represent this operator in position basis we
define a  block matrix of $n^{\levelq}$ Fourier transformations,
\begin{equation}
    \ufourier =  \diag( \mathcal{F}_{N/n^{\levelq}}, \dots, \mathcal{F}_{N/n^{\levelq}}),
\end{equation}
acting on vertical phase-space regions that consist each of $n^{\levelp}$ randomization regions.
The randomization operator in
position basis is then given by
\begin{equation}
    \urandom = \ufourier^{-1}\ 
    \hat{\mathcal{U}}^{\mathrm{cue}}_{\levelp,\levelq}\
    \ufourier.
\end{equation}
Altogether, we define the randomized baker map with escape as the composition of randomization, escape and baker map,
\begin{equation}
    \qbmaprrandom
    := \qbmap\, \qrefl\, \urandom.
    \label{eq:def-qbmaprandom}
\end{equation}

\subsection{Resonance states}
\label{sec:resonance-states}
Right resonance states of the randomized baker map with escape are defined
by
\begin{equation}
    \qbmaprrandom\  \psi = \ue^{\ui \theta - \gamma/2}\ \psi,
\end{equation}
where $\gamma$ is the decay rate, i.e., the
square of the norm of $\psi$ decays by a factor of $\ue^{-\gamma}$ when $\qbmaprrandom$ is applied.
The phase $\theta$ and decay rate $\gamma$ are
related to the energy and width of resonance
poles in autonomous scattering \cite{FyoSom1997,Sch2013b}.
The corresponding left resonance states are not discussed separately, 
as one can show that they are right resonance states 
for the related operator, where
$\qrefl$ is replaced with $\qrefl^{-1}$ in Eq.~\eqref{eq:def-qbmaprandom},
i.e., with inverted reflectivity constants, $1 / \refl_k$.

As we are interested in the phase-space distribution of
resonance states $\psi$,
we consider the Husimi function $\hus[{\psi}]$ \cite{Hus1940,NonVor1998}
which is a smooth probability density.
It is defined using the overlap of the state $\psi$ with a
coherent state $c(q, p)$ centered at some phase-space
point $(q,p)$,
\begin{equation}
    \hus[\psi](q,p) = N \; \Vert 
    \langle{c(q,p)}|{\psi}\rangle \Vert^2,
\end{equation}
where $N$ is the Hilbert space dimension.
For quantum maps on the torus the coherent states
are defined in Ref.~\cite{NonVor1998}.
In the following numerical illustrations the width of
the coherent states is chosen to be symmetric in phase space.

\begin{figure}[t!]
    \centering
\includegraphics[scale=1]{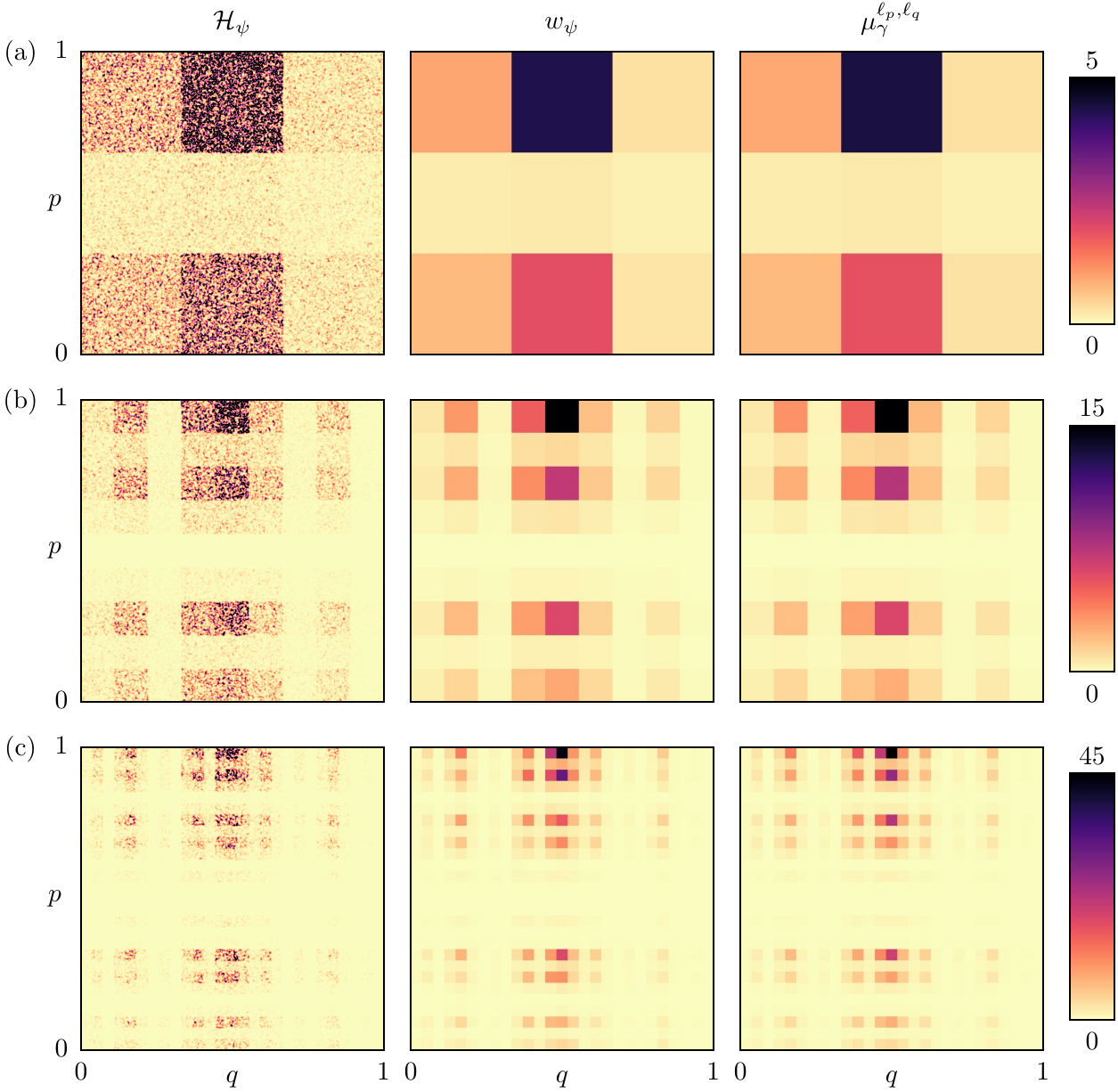}
    \caption{Phase-space distributions of resonance states
             and classical measures
             for randomization regions on increasingly finer levels
             (a) $(\levelp, \levelq) = (0, 1)$,
             (b) $(\levelp, \levelq) = (1, 2)$, and 
             (c) $(\levelp, \levelq) = (2, 3)$
             for the randomized $3$-baker map with $\rbold = (0.2, 0.01, 1)$.
     Shown is (left) Husimi function $\hus$ for the resonance state
     $\psi$ with decay rate $\gamma$ closest to $\gtyp$ 
     for $N=25515$, 
     (middle)
     projected weights 
     $w_\psi([\brecta_{\levelp + 1, \levelq}])$ 
     on the level of the randomization subregions,
     and
     (right)
     corresponding classical measure $\mulevel$
     derived in Sec.~\ref{sec:semiclassics}.
     }
    \label{fig:husimi}
\end{figure}

We illustrate in Fig.~\ref{fig:husimi}(a) in the left panel a resonance state 
for the randomized $3$-baker map with $n=3$ stripes, reflectivities
$\boldsymbol{\refl} = (0.2, 0.01, 1)$ and $N = 35\cdot 3^6 = 25515$
with decay rate closest to $\gtyp$.
Here the randomization regions are defined by
$(\levelp, \levelq)  = (0, 1)$, i.e., they correspond to the rectangles
$A_k$.
One observes that the resonance state is not 
uniformly fluctuating in each of the $n=3$ randomization regions.
Instead, we find a finer substructure,
where each randomization region has $n=3$ subregions in $p$-direction,
leading to a $3\times 3$ grid.
This subregion structure is discussed in more detail in \ref{app:subregion-structure}.

Within each subregion the resonance state visually fluctuates as in
closed chaotic quantum maps and we find an
exponential distribution around the
average, as conjectured in Ref.~\cite{ClaKunBaeKet2021}
for quantum maps with escape.
The subregion structure remains the same for increasing
$N$ and the Planck cell becomes arbitrarily small compared to it,
giving a well-defined semiclassical limit.
This is in contrast
to deterministic quantum maps with escape, where even for arbitrarily large
$N$ the phase-space structure of a resonance state is multifractal down to
the scale of a Planck cell.

In Figs.~\ref{fig:husimi}(b) and (c) the randomization regions are defined
on finer scales by $(\levelp, \levelq)  = (1, 2)$ and $(\levelp, \levelq)  = (2, 3)$,
respectively.
The resonance states (left panel) are again uniformly fluctuating on $n$ subregions
of each randomization region, leading to a
$n^{\levelp + 1} \times n^{\levelq}$-grid structure.
Again, for increasing $N$ the Planck cell size becomes arbitrarily small compared
to the phase-space structures of resonance states.

For increasingly finer randomization levels we observe in
Figs.~\ref{fig:husimi}(a)--(c)
that the phase-space structure on smaller scales resembles
(qualitatively) the phase-space structure on larger scales.
For example the region $(q, p) \in [1/3, 2/3] \times [2/3, 1]$
in Fig.~\ref{fig:husimi}(b) looks similar to the full phase space
in Fig.~\ref{fig:husimi}(a).
In the limit of $\levelp,\levelq \rightarrow \infty$ with
sufficiently large $N$ this leads to multifractal structures.

Note as an aside, that if for a fixed Hilbert space dimension $N$,
the number $n^{\levelp + \levelq}$ of randomization regions is increased,
the dimension $\ncue = N /  n^{\levelp + \levelq}$ of the random matrices becomes smaller
as well as
the number of Planck cells 
$\nsub = \ncue / n = N /  n^{\levelp + 1 + \levelq}$ within each randomization subregion.
In Figs.~\ref{fig:husimi}(a)--(c) this leads to
$\nsub = 2835$,
$\nsub = 315$, and
$\nsub = 35$, respectively.
Numerically, this limits the number of levels that can be explored, while still
having $\nsub \gg 1$.

\subsection{Weight of resonance state on rectangular phase-space regions}

In order to compare the phase-space structure of a resonance state
with some classical measure, 
we choose rectangular phase-space regions onto which we project the
resonance state.
A suitable projector is defined by first projecting the position representation to
the interval corresponding to the extent of the rectangle in $q$-direction.
The second step is a transformation to the momentum representation and 
projection on the interval corresponding to the extent of the rectangle in $p$-direction.
For convenience we use the rectangles $[\brecta_{m_p, m_q}]$
partitioning the phase space based on the finite symbolic sequences of the baker map.
We define a projector $P_{[m_p, m_q]}$
for all $n^{m_p + m_q}$ possible finite sequences $\brecta_{m_p, m_q}$ 
with $\alpha_i \in \{0, \dots, n -1\}$, 
\begin{equation}
	\hspace*{-1.5cm}
	P_{[\brecta_{m_p, m_q}]} =
	\ufourier[m_q]^{-1} 
	\;
	\diag(\underbrace{\mathbb{O}_{\nrect}, \dots, \mathbb{O}_{\nrect}}_{
		I(\brecta_{m_p, m_q})}  
	, \mathds{1}_{\nrect},
	\underbrace{\mathbb{O}_{\nrect}, \dots, \mathbb{O}_{\nrect} }_{n^{m_p + m_q} -   I(\brecta_{m_p, m_q}) - 1})\ 
	\;
	\ufourier[m_q],
\end{equation}
where $\nrect$ is the dimension of the Hilbert space of each rectangle
and $N = n^{m_p + m_q} \nrect$ 
the total Hilbert space dimension. 
Here the rectangles $[\brecta_{m_p, m_q}]$ are ordered according to the index
$I(\brecta_{m_p, m_q})
= \sum_{i=0}^{m_p - 1}  \alpha_{i - m_p} n^i
+ n^{m_p}
\sum_{i=0}^{m_q - 1} \alpha_{m_q - 1 - i} n^i$,
which counts the rectangles first in the leftmost column from
bottom to top,
then in the second column, and so on.
The projected weight of a state $\psi$ on a rectangle $[\brecta_{m_p, m_q}]$ is then given by
\begin{equation}
	w_\psi([\brecta_{m_p, m_q}])
	= \Vert P_{[\brecta_{m_p, m_q}]} \psi \Vert^2.
    \label{eq:def-projected-weights}
\end{equation}

Alternatively, one may determine the weights by integrating the Husimi function over the corresponding
rectangle on $\torus$,
$w_\psi([\brecta_{m_p, m_q}])
= \int_{{J_n}({[\brecta_{m_p, m_q}]})} \hus[\psi](q,p) \, \ud \muL$.
Note that the weights of both approaches
agree in the semiclassical limit $N\rightarrow \infty$.
Convergence is slower when using the Husimi function, as the weight in rectangles with small weights 
is strongly influenced by neighboring rectangles with large weights due to the finite resolution of the Husimi function. 

In Figs.~\ref{fig:husimi}(a-c) in the middle panel
the weight on all subregion rectangles $[\brecta_{\levelp + 1, \levelq}]$
of a resonance state is illustrated. 
This can be compared to classical measures in the right panel,
derived in Sec.~\ref{sec:semiclassics},
showing excellent agreement on a qualitative level.
A quantitative comparison will be presented in Sec.~\ref{sec:numerical}.
We emphasize that while Fig.~\ref{fig:husimi} shows results for just one 
exemplary decay rate $\gtyp$,
the weight in each subregion rectangle depends on the decay rate $\gamma$.
It is the goal of this paper to derive classical measures
for arbitrary decay rates.
In the next section we consider randomization regions 
with arbitrary but fixed $\levelp$ and $\levelq$ 
and make a conjecture for the 
semiclassical limit $N \rightarrow \infty$.

\section{Semiclassical structure of resonance states}
\label{sec:semiclassics}

In this section we present a conjecture for the semiclassical
structure of resonance states of the 
randomized baker map with partial escape.
To this end we introduce classical measures, which 
(i) are conditionally invariant
with some desired decay rate $\gamma$
on scales larger than the scale of randomization
and (ii) obey an extremum principle derived from a
local random vector model in each randomization region.

\subsection{Conditional invariance on rectangles}
\label{sec:cinv}

Semiclassically, resonance states of quantum maps with escape
converge to conditionally invariant measures of the same decay rate \cite{KeaNovPraSie2006,NonRub2007,ClaKoeBaeKet2018}.
A measure $\mu$ on $\torus$ is conditionally
invariant \cite{DemYou2006} with respect to the classical baker map
$\bmap$ with escape $r$,
if there exists a decay rate $\gamma \in \mathbb{R}$ with
\begin{equation}
    \int_{\bmap^{-1}(A)} \refl\, \ud\mu  =
    \ue^{-\gamma} \, \mu(A) 
    \qquad  \forall A \subset \torus.
    \label{eq:def-cinv}
\end{equation}
For constant reflectivity factors $\refl_k$ from the rectangles $A_k$, as considered in
this paper, this condition simplifies for subsets of 
$A_k$ to
\begin{equation}
 \refl_k \, \mu(C) =
 \ue^{-\gamma} \, \mu(\bmap(C))
 \qquad  \forall C \subset A_k.
 \label{eq:ci-constant}
\end{equation}
In general, \EQref{eq:def-cinv} holds for all measurable
sets $A\subset \torus$ up to arbitrary small scales,
which is related to the multifractal structure of 
conditionally invariant measures.
However, for the randomized quantum baker map, 
as introduced in the last section,
we observe no structure of resonance states on scales finer than
the subregions of the randomization regions
$[\brecta_{\levelp, \levelq}]$,
see Figs.~\ref{fig:husimi}(a-c) left panel.
For this reason, conditional invariance cannot be satisfied
on arbitrarily fine scales.
This motivates to introduce the following scale dependent definition.
We call a measure $\mu$ on $\torus$ conditionally invariant 
down to the scale $(\levelp, \levelq)$ with respect
to the baker map with escape $\bmap[n,\rbold]$,
if Eq.~\eqref{eq:ci-constant} holds for all sets
$C = J_n([\brecta_{\levelp,\levelq}])$.

We assume for the randomized quantum baker map,
that the resonance states 
converge in the semiclassical limit
to measures which satisfy conditional invariance
down to the scale $(\levelp, \levelq)$ of the
randomization regions
and are constant on the subregions of scale $(\levelp + 1, \levelq)$.
Such measures are shown in 
Figs.~\ref{fig:husimi}(a-c) right panel
and are derived in this section.

From conditional invariance down to the scale of randomization  we derive
constraints for $\mu$.
Recall that the randomization regions are defined as the 
rectangles
$[\brecta_{\levelp, \levelq}]
=
[\recta_{-\levelp} \dots \recta_{-1}\cdot%
\recta_0 \recta_1 \dots \recta_{\levelq - 1}]$
with $\alpha_i \in \{0, \dots, n-1\}$ for all
$-\levelp \leq i \leq \levelq - 1$.
Their iterates are given by
$[\bshift{}(\brecta_{\levelp, \levelq})]
= [\recta_{-\levelp} \dots \recta_{-1}%
\recta_0 \cdot \recta_1 \dots \recta_{\levelq - 1}]$.
These iterates correspond to rectangles, which are 
stretched by a factor $n$ in $q$-direction and compressed
by $1/n$ in $p$-direction.
Each of these iterates overlaps with exactly
$n$ of the randomization regions (and vice versa),
see illustration in Fig.~\ref{fig:sketch}(b).
Hence, the original and the iterated randomization rectangles 
can be partitioned as follows,
\begin{eqnarray}
   [\brecta_{\levelp, \levelq}] 
    &= \bigcup_{j = 0}^{n-1} [j\brecta_{\levelp, \levelq}],
    \qquad %
   [\bshift{} (\brecta_{\levelp, \levelq})] 
= \bigcup_{j = 0}^{n-1} [\bshift{}(\brecta_{\levelp, \levelq}) \, j]
\label{eq:partition}
\end{eqnarray}
where for each $j\in\{0, \dots, n -1\}$ the smaller rectangles
$[j\brecta_{\levelp, \levelq}]:=
[j\,\recta_{-\levelp} \dots \recta_{-1} \cdot%
 \recta_0 \recta_1 \dots \recta_{\levelq - 1}]
$
and
$[\bshift{}(\brecta_{\levelp, \levelq}) \, j]:=
[\recta_{-\levelp} \dots \recta_{0} \cdot%
\recta_1 \recta_2 \dots \recta_{\levelq - 1} \, j]
$
are the subregions of the randomization regions
on which the measure $\mu$ is constant.

Conditional invariance down to the scale of the randomization regions
$[\brecta_{\levelp, \levelq}]$ 
implies, by setting $C = J_n([\brecta_{\levelp, \levelq}])$
in Eq.~\eqref{eq:ci-constant} with
$C \subset A_{\alpha_0}$, that
\begin{equation}
     \refl_{\recta_0} \, \mu([\brecta_{\levelp, \levelq}])
    =
    \ue^{-\gamma} \, \mu([\bshift{}(\brecta_{\levelp, \levelq})])
\end{equation}
for all multi-indices $\brecta_{\levelp, \levelq}$.
Inserting the
partitions from Eq.~\eqref{eq:partition} we get
\begin{equation}
\refl_{\recta_0} \,
\mu\left(\bigcup_{j = 0}^{n-1} [j\brecta_{\levelp, \levelq}]
\right)
=
\ue^{-\gamma} \, 
\mu\left(\bigcup_{j = 0}^{n-1} [\bshift{}(\brecta_{\levelp, \levelq}) \, j]
\right).
\end{equation}
This gives the constraints of conditional invariance on
the weights of all subregion rectangles with scales
$(\levelp + 1, \levelq)$,
\begin{eqnarray}
    \refl_{\recta_0}
    \sum_{j = 0}^{n-1}\mu([j \brecta_{\levelp, \levelq}])
    =
    \ue^{-\gamma} 
    \sum_{j = 0}^{n-1} \mu([\bshift{}(\brecta_{\levelp, \levelq}) \, j]),
    \label{eq:constraints-ci}
\end{eqnarray}
holding for all possible finite symbolic sequences of the form
$\brecta_{\levelp, \levelq} =
\recta_{-\levelp}\dots\recta_{-1}.\recta_0\dots\recta_{\levelq-1}$ with $\recta_i \in \{0, \dots, n-1\}$.
Altogether, these are $n^{\levelp + \levelq}$
constraints on the subregion weights.

Normalization of the measure $\mu$ 
\begin{eqnarray}
    \mu(\torus) 
    &=
    \mu\left(
    \bigcup_{j = 0}^{n-1}
    \bigcup_{\recta_{-\levelp} = 0}^{n-1}
    \dots
    \bigcup_{\recta_0 = 0}^{n-1}
    \dots
    \bigcup_{\recta_{\levelq - 1} = 0}^{n-1}
    [j\brecta_{\levelp, \levelq}]
    \right)
    = 1
\end{eqnarray}
adds one more
constraint 
\begin{eqnarray}
	\sum_{j = 0}^{n-1}
	\sum_{\recta_{-\levelp} = 0}^{n-1}
	\dots
	\sum_{\recta_0 = 0}^{n-1}
	\dots
	\sum_{\recta_{\levelq - 1} = 0}^{n-1}
	\mu([j\brecta_{\levelp, \levelq}]) 
	= 1.
	\label{eq:constraints-norm}
\end{eqnarray}
Thus, conditional invariance down to scale $(\levelp, \levelq)$, Eq.~\eqref{eq:constraints-ci}, 
and normalization, Eq.~\eqref{eq:constraints-norm},
give $n^{\levelp + \levelq} + 1$
constraints on the weights of all
$n^{\levelp + 1 + \levelq}$ subregion rectangles.
This leads to the question, how
quantum mechanics
distributes the weight over the 
subregion rectangles under the restrictions of
Eqs.~\eqref{eq:constraints-ci} and
\eqref{eq:constraints-norm}.
This will be answered in the next section.

\subsection{Local random vector model}
\label{sec:local-random-wave}

Based on a random vector model \cite{NonVor1998,BaeNon:p}
that is applied locally in every randomization region, 
we derive in the following the semiclassical structure
of resonance states of the randomized quantum baker map
and its dependence on the decay rate $\gamma$.
We introduce the local random vector model
for each randomization region as follows.
Let $[\brecta_{\levelp,\levelq}]$ be one of these randomization regions
with $\ncue$ sites,
corresponding to the number of
Planck cells in this randomization region.
Let site weight factors $f_i \ge 0$ with $\sum_{i=0}^{\ncue - 1} f_i = 1$
describe how
the a priori unknown weight $\mu([\brecta_{\levelp,\levelq}])$
of the randomization region is distributed on its sites,
such that the weight at site $i$ is given by $f_i \, \mu([\brecta_{\levelp,\levelq}])$.
Let these site weight factors $f_i$ be given by the intensities 
$f_i = |c_i|^2$ of a random vector with complex amplitudes $c_i \in \mathbb{C}$,
which must be normalized 
$\sum_{i=0}^{\ncue - 1} |c_i|^2 = 1$,
following the definition of the factors $f_i$.

For such normalized random vectors of dimension $\ncue$ the joint probability distribution
of its complex amplitudes is given by \cite{BroFloFreMelPanWon1981, KusMosHaa1988},
\begin{equation}
	P(\{c_i\})
	= 
	\frac{\Gamma(\ncue)}{\pi^{\ncue}}
	\;
	\delta \! \left( \sum_{i=0}^{\ncue-1} |c_i|^2  - 1 \right)
	.
\end{equation}
By integration over the angle of each complex amplitude
it follows that its intensities, i.e., the site weight factors $f_i = |c_i|^2$,
have the joint probability distribution
\begin{equation}
	P(\{f_i\})
	= 
	\Gamma(\ncue)
	\;
	\delta \! \left( \sum_{i=0}^{\ncue-1} f_i - 1 \right)
	,
\label{eq:f-factor}
\end{equation}
which is uniform for normalized factors $f_i$. This uniformity will be used below.
In the special case that
the randomization region is the entire phase space $\torus$,
which has measure $\mu(\torus) = 1$,
this is equivalent to the usual
random vector model
for chaotic maps \cite{NonVor1998,BaeNon:p}.

Each randomization region consists of $n$ subregions 
$[j\brecta_{\levelp,\levelq}]$ 
with $j\in\{0, \dots,  n-1\}$,
where each subregion has $\nsub$ sites
if $\ncue = n \nsub$ is chosen appropriately.  
We define a subregion weight factor $F_j$ by
the relative weight of the set $[j\brecta_{\levelp,\levelq}]$ to $[\brecta_{\levelp,\levelq}]$,
\begin{equation}
F_j = \frac{\mu([j\brecta_{\levelp,\levelq}])}{\mu([\brecta_{\levelp,\levelq}])}.
\label{eq:subregion-weights0}
\end{equation}
The $F_j$ are thus
given by the sum over all corresponding
site weight factors $f_i$.
For a suitable ordering of the $f_i$ this can be expressed as
\begin{equation}
\label{eq:subregion_weights}	
F_j = \sum_{i = 0}^{\nsub-1} f_{j\nsub + i}
\qquad
\forall j=0,\dots,n-1
.
\label{eq:subregion-wfactor}
\end{equation}
Note that normalization of the subregion weight factors,
$\sum_{j=0}^{n-1} F_j = \sum_{i=0}^{\ncue-1} f_i = 1$, 
follows from the above normalization of the 
site weight factors $f_i$.

The joint probability distribution $P(\{F_j\})$ 
of subregion weight factors in a randomization region 
is given by integrating over the
uniform joint probability distribution of site weight factors,
Eq.~(\ref{eq:f-factor}), while
fulfilling Eqs.~\eqref{eq:subregion_weights} for all $j$,
\begin{eqnarray}
\hspace*{-2.3cm}P(\{F_j\}) 
    &=&
    \int \prod_{j=0}^{n-1}
    \delta \! \left( F_j - \sum_{i=0}^{\nsub - 1} f_{j\nsub + i} \right)
                       \  \ud P(\{f_i\}) 
    \\
\hspace*{-2.3cm} &=&
    \Gamma(\ncue)\;\delta\!\left(\sum_{j=0}^{n-1}F_j -1 \right)
	\prod_{j=0}^{n-1}
	\int \delta \! \left( F_j - \sum_{i=0}^{\nsub - 1} f_{j\nsub + i} \right)
    \ \prod_{i=0}^{\nsub -1}\ud f_{j\nsub + i}.
    \nonumber
\end{eqnarray}
Here for each $j$ the integral gives the volume of
an $(\nsub - 1)$-dimensional standard simplex within
the $\nsub$-dimensional space $\{f_{j\nsub}, f_{j\nsub + 1}, \dots, f_{j\nsub + \nsub - 1}\}$, which is
scaled in all dimensions with the factor $F_j$
and thus is proportional to $F_j^{\nsub - 1}$.
For example, for $\nsub = 3$ it is a triangle in the 3D space
$(f_{3j},f_{3j+1},f_{3j+2})$
with corners $(F_j,0,0)$, $(0,F_j,0)$, $(0,0,F_j)$
and 2D volume proportional to $F_j^2$.
Thus we obtain for the joint probability distribution  of normalized subregion weight factors,
\begin{equation}
	P(\{F_j\}) \propto
	\prod_{j=0}^{n-1} F_j^{\nsub - 1}
	.
\end{equation}

We now extend this argument to all $n^{\levelp + \levelq}$ randomization regions
$[\brecta_{\levelp,\levelq}]$, each with $n$
subregion weight factors $F_{j\brecta_{\levelp,\levelq}}$ from the local random vector model.
Assuming independence,
the joint probability distribution of all subregion weight factors is proportional to
\begin{equation}
	P(\{ F_{j\brecta_{\levelp,\levelq}} \}) \propto
	\prod_{\brecta_{\levelp,\levelq}} \prod_{j=0}^{n-1} 
	F_{j\brecta_{\levelp,\levelq}}^{\nsub - 1}
	=
	\Bigg(
	\prod_{\brecta_{\levelp,\levelq}} \prod_{j=0}^{n-1} 
	F_{j\brecta_{\levelp,\levelq}}
	\Bigg)^{\nsub - 1}.
\end{equation}
In the semiclassical limit, $\nsub \rightarrow \infty$,
this distribution approaches a delta function at the maximum of
the product of the subregion weight factors,
$\prod_{\brecta_{\levelp,\levelq}} \prod_{j=0}^{n-1} 
F_{j\brecta_{\levelp,\levelq}}$.
In this product we insert the definition of the subregion weight factors,
Eq.~\eqref{eq:subregion-weights0},
\begin{equation}
	F_{j\brecta_{\levelp,\levelq}}
	=
	\frac{\mu([j\brecta_{\levelp,\levelq}])}{\sum\limits_{l=0}^{n-1}\mu([l\brecta_{\levelp,\levelq}])}
	,
\end{equation}
using the subregion weights
$\mu([j\brecta_{\levelp,\levelq}])$
and the weight of the corresponding randomization region
$\mu([\brecta_{\levelp,\levelq}]) = \sum_{l=0}^{n-1}\mu([l\brecta_{\levelp,\levelq}])$.
This allows for writing the product of subregion weight factors
as a product of relative weights,
\begin{equation}
	\Pi(\mu)
	=
	\prod_{\brecta_{\levelp,\levelq}} \prod_{j=0}^{n-1}
	\frac{\mu([j\brecta_{\levelp,\levelq}])}{\sum\limits_{l=0}^{n-1}\mu([l\brecta_{\levelp,\levelq}]) }
	.
	\label{eq:product-lrwm}
\end{equation}

Thus the local random vector model for each randomization region leads to
the following extremum principle:
The subregion weights $\mu({[j\brecta_{\levelp,\levelq}]})$ maximize the product $\Pi(\mu)$
of relative weights, Eq.~\eqref{eq:product-lrwm},
under the constraints of conditional invariance, Eq.~\eqref{eq:constraints-ci},
and normalization, Eq.~\eqref{eq:constraints-norm}.
They can be determined, e.g.,
by using the method of Lagrange multipliers 
leading to a set of coupled nonlinear equations.
These subregion weights $\mu([j\brecta_{\levelp,\levelq}])$
define a classical measure $\mulevel$ on $\torus$,
that is conditionally invariant
down to the scale of the randomization regions, determined
by $\levelp$ and $\levelq$,
and that depends on the decay rate $\gamma$.

In Figs.~\ref{fig:husimi}(a-c) in the right panel
examples of such classical measures $\mulevel$
for three increasing levels of randomization
$\levelp \in \{0, 1, 2\}$ with $\levelq = \levelp + 1$
are shown.
We find quite good agreement with the structure of the resonance states (left panel)
and the projected weights within each subregion rectangle (middle panel).
Slight deviations are observable in (c), where 
the subregion dimension $\nsub=35$ is quite small and the projected weights
are still influenced by the fluctuations
of an individual resonance state.

\subsection{Semiclassical conjecture}
\label{sec:semiclassics-conjecture}

The following conjecture is formulated in
analogy to quantum ergodicity in closed chaotic maps
\cite{NonVor1998,MarKeeZel2005}.
For fixed $\levelp$ and $\levelq$ we conjecture that
the classical measures $\mulevel$ are the semiclassical limit
measures of resonance states $\psi$ of the randomized quantum baker
map with escape $\qbmaprrandom $,
which decay with the rate~$\gamma$.
In particular, let
$\{\qbmaprrandom^N\}$ be the ensemble of
randomized quantum baker maps with dimension $N$.
Let $\psi_N$ be an eigenstate of any member $\qbmaprrandom^{N}$ of the ensemble,
$\qbmaprrandom^{N}\psi_N = \lambda_N\psi_N$.
For a sequence of such states with decay rates $\gamma_N = -\ln |\lambda_N|^2 \overset{N\rightarrow\infty}{\longrightarrow}\gamma$,
and for all observables $a$ on $\torus$ with quantization
$\op(a)$ we expect
\begin{equation}
    \langle \psi_N | \op(a) |\psi_N \rangle
    \overset{N\rightarrow\infty}{\longrightarrow}
     \mulevel(a) = \int_\torus a \, \ud\mulevel.
	\label{eq:semiclassical-conjecture}
\end{equation}
This means, that the quantum expectation values converge
to the classical ones given by $\mulevel$.
Note that there are exceptional sequences $\psi_N$ from exceptional members
of the ensemble $\{\qbmaprrandom^N\}$,
such that convergence in Eq.~\eqref{eq:semiclassical-conjecture}
is expected just for a suitable subsequence of density one.
A quantitative analysis of the quantum-classical agreement
is given in Sec.~\ref{sec:numerical}.

\subsection{Properties of the classical measure $\mulevel$}
\label{sec:properties}

We list some relevant properties and remarks on the
classical measures $\mulevel$:

\paragraph{(i)}
Numerics for increasing $\levelp, \levelq$ 
suggests that in the limit $\levelp, \levelq\rightarrow \infty$
the classical measures $\mulevel$ converge weakly to some limiting
measure $\mu_\gamma^\ast$, see \ref{app:convergence-measures}.
\paragraph{(ii)}
The product $\Pi(\mu)$
of relative weights, Eq.~\eqref{eq:product-lrwm},
is maximal if all relative weights are equally given by $1/n$.
Assuming conditional invariance one can show that this implies that the measures are also uniform within the sets
$[\brecta_{0,\levelq}]$, i.e., along the $p$-direction.
The subregion weights follow as
$\mu({[\brecta_{\levelp + 1,\levelq}]}) 
= \prod_{i=0}^{\levelq - 1} \refl_{\recta_i}^{-1} / \mathcal{N}$
with some normalization $\mathcal{N}$.
This measure is conditionally invariant with decay rate $\ginv$,
the natural growth rate of the inverse map, see Sec.~\ref{sec:classical-baker-escape}.
\paragraph{(iii)}
For the natural decay rate $\gnat$ one needs to rewrite the product $\Pi(\mu)$
using conditional invariance, Eq.~\eqref{eq:constraints-ci}, 
with decay rate $\gnat$, leading after reordering to
\begin{equation}
	\Pi(\mu)
	=
	\prod_{\brecta_{\levelp,\levelq}} \prod_{j=0}^{n-1}
	\refl_{\recta_{-1}} \ue^{\gnat}
	\frac{\mu([\brecta_{\levelp,\levelq}j])}{\sum\limits_{l=0}^{n-1}\mu([\brecta_{\levelp,\levelq}l]) }
	.
	\label{eq:product-lrwm-nat}
\end{equation}
This product is maximal if all relative weights, now with respect to iterated
randomization regions, are equally given by $1/n$.
One can show that this implies that all
subregion weights are given by
$\mu({[\brecta_{\levelp + 1,\levelq}]}) 
= \prod_{i=-1}^{-\levelp - 1} \refl_{\recta_{i}} / \mathcal{N}$
with some normalization $\mathcal{N}$,
i.e., the measure is uniform in $q$-direction.
This measure is indeed conditionally invariant with decay rate $\gnat$.

\paragraph{(iv)}
The classical measure $\mulevel$ for the special case $\gnat$ ($\ginv$)
is consistent with the general observation that
at the (inverse) natural decay rate resonance states are uniform 
along the unstable (stable) direction \cite{AltPorTel2013,ClaAltBaeKet2019}.

\paragraph{(v)}
We mention as an aside, that
if one applies this extremum principle to closed chaotic systems,
it leads to the uniform distribution, which
is consistent with quantum ergodicity
\cite{DegGraIso1995,BouDeB1996,NonVor1998}.

\paragraph{(vi)}
It is possible to apply the local random vector model
combined with conditional invariance
to random matrices with partial escape,
consistently retrieving the results of Ref.~\cite[App.~D]{ClaKunBaeKet2021}.
The details are given in \ref{sec:rmt}.

\paragraph{(vii)}
    If simultaneously decreasing the scale of the randomization regions and increasing
    $N$,
    such that still $\nsub \rightarrow \infty$ holds,
    we expect that the local random vector model leads to the same limiting measures
    $\mu_\gamma^\ast$, as discussed in remark \textit{(i)} above.

\section{Numerical verification}
\label{sec:numerical}

In this section we present quantitative numerical support for 
the semiclassical conjecture given in the last section,
going beyond the qualitative comparison of
Fig.~\ref{fig:husimi}.
All results are presented for single realizations
of the randomized quantum baker map without averaging.

\begin{figure}[t!]
    \centering
    \includegraphics[scale=1]{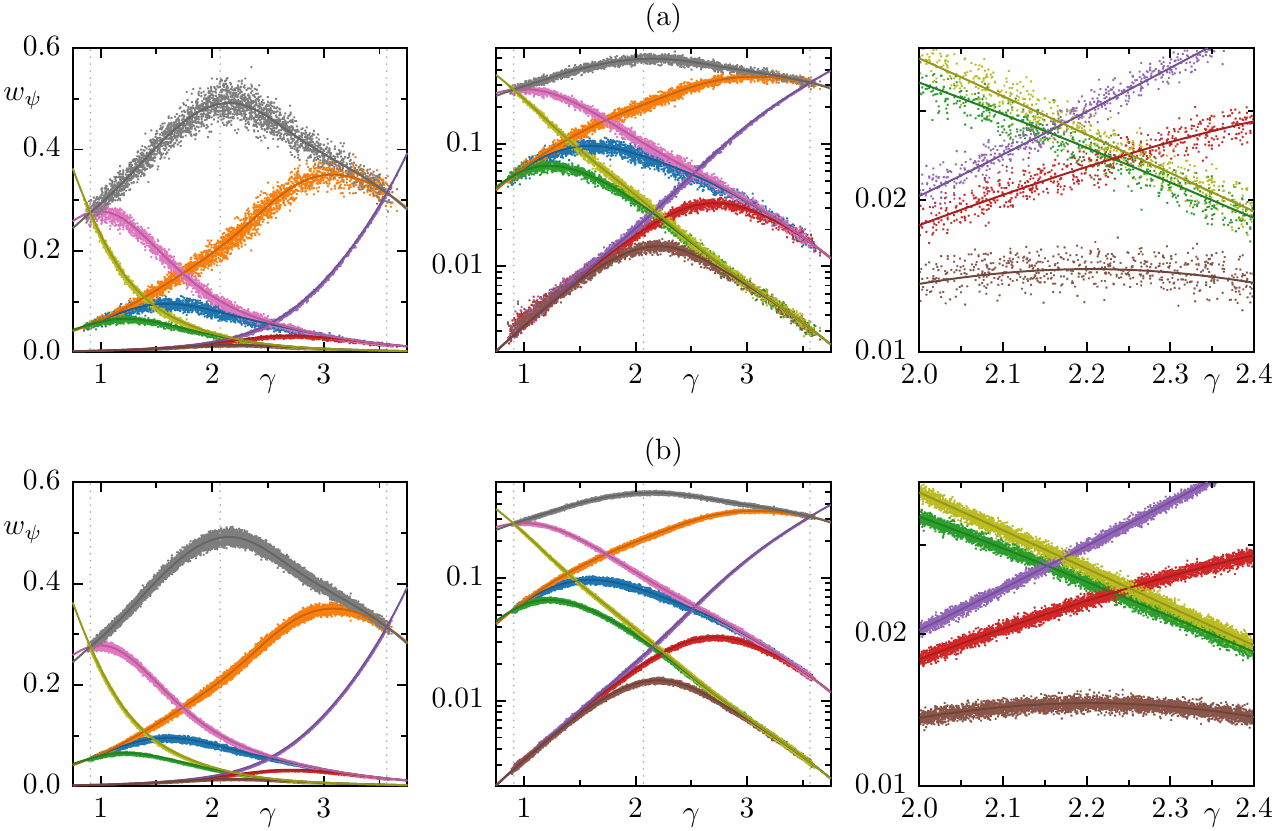}
    \caption{Comparison of the 9 projected weights
        $w_{\psi_i}([\brecta_{1,1}])$
        of each resonance state $\psi_i$
        vs.\ its decay rate $\gamma_i$ (dots)
        and classical measures
        $\mulevel([\brecta_{1,1}])$ (lines)
        for $3$-baker map with $\rbold = (0.2, 0.01, 1)$
        and randomization level $(\levelp, \levelq) = (0, 1)$
        for Hilbert space dimension (a) $N = 2835$ and (b) $N = 25515$.
        The same data is shown on
        linear (left) and logarithmic scale (middle) 
        as well as a magnification (right).
        Decay rates $\gnat$, $\gtyp$, and $\ginv$ are indicated (dotted lines).
    }
    \label{fig:weights_n31}
\end{figure}
\subsection{Weights on classically large regions}
In order to test the semiclassical conjecture, 
Eq.~\eqref{eq:semiclassical-conjecture},
we choose as observables the projection operators on classically 
large regions compared to a Planck cell.
Specifically, we use squares given by 
symbolic sequences
$[\brecta_{\level,\level}]$ with $\level = 1$
and $\level = 2$.
For different systems and Hilbert space dimensions
we consider all resonance states
$\psi_i$ and show how the projected weights 
$w_{\psi_i}([\brecta_{\level, \level}])$, Eq.~\eqref{eq:def-projected-weights},
depend on their quantum decay rate $\gamma_i$.
This is compared to the classical measures $\mulevel([\brecta_{\level, \level}])$
on these squares and their dependence on $\gamma$.
The Hilbert space dimension $N$ is in all cases chosen to be a multiple of 
the number $n^{\levelp + 1 + \levelq}$ of subregions.

In Fig.~\ref{fig:weights_n31}
we consider the $3$-baker map with $\rbold = (0.2, 0.01, 1)$
and randomization level $(\levelp, \levelq) = (0, 1)$
for increasing Hilbert space dimension.
For $\level = 1$ there are $n^{2 \level} = 9$
quantum and classical weights which are compared.
We observe that the quantum weights (dots) scatter symmetrically around
the classical measures (lines). 
The width of the distribution decreases with increasing Hilbert space dimension 
for each of the considered squares
(compare (a) and (b)), supporting the semiclassical conjecture.
This also holds for small weights, as can be seen on the 
logarithmic scale (middle panel).

At the natural decay rates $\gnat$ and $\ginv$
the measures are constant along one direction, see Sec.~\ref{sec:properties},
and several weights are identical
in agreement with the properties of the resonance states.

\begin{figure}[t!]
    \centering
    \includegraphics[scale=1]{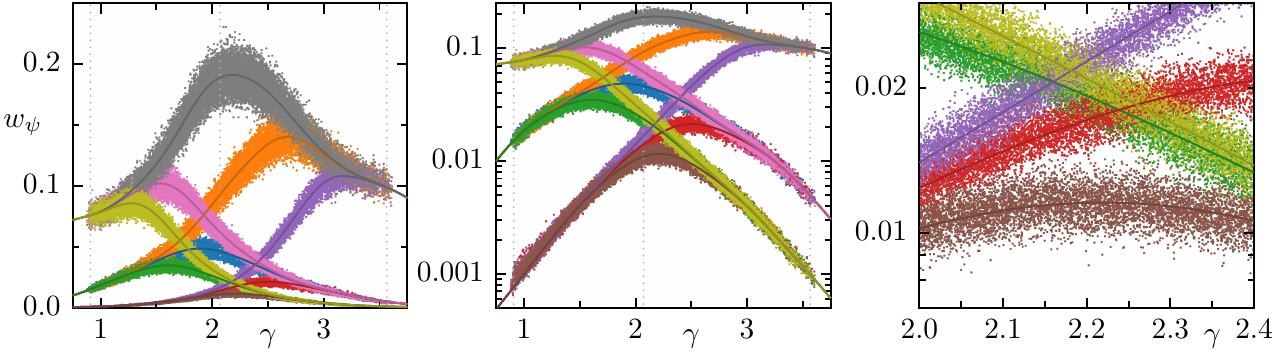}
    \caption{%
        Same as Fig.~\ref{fig:weights_n31}(b), 
        but for randomization level $(\levelp, \levelq) = (1, 2)$ 
        and for 9 selected squares from $[\brecta_{2,2}]$
        with symbolic sequence $\brecta = i2 \cdot 1j$ and $i, j\in \{0, 1, 2\}$.
    }
    \label{fig:weights_n32}
\end{figure}

Figure~\ref{fig:weights_n32} shows the comparison for smaller
randomization regions
$(\levelp, \levelq) = (1, 2)$
and
weights on smaller phase-space regions 
$[\brecta_{\level,\level}]$ with $\level = 2$.
There are $n^{2 \level} = 81$ such regions out of which
9 are presented in the figure,
corresponding to the phase-space region 
$(q, p) \in [1/3, 2/3] \times [2/3, 1]$,
see Fig.~\ref{fig:husimi}(a).
Again, quantum and classical weights show excellent agreement.
The width of each distribution is wider than in 
Fig.~\ref{fig:weights_n31}(b), 
as the observables are smaller phase-space regions.
For larger Hilbert space dimension $N$ it is expected to decrease.
We emphasize that the dependence on $\gamma$ qualitatively resembles
the results for $(\levelp, \levelq) = (0,1)$ shown in Fig.~\ref{fig:weights_n31},
but now on a smaller scale in phase space. 
This indicates again the multifractal phase-space structure, 
which emerges for smaller randomization regions,
as observed in Fig.~\ref{fig:husimi}.

\begin{figure}[t!]
    \centering
    \includegraphics[scale=1]{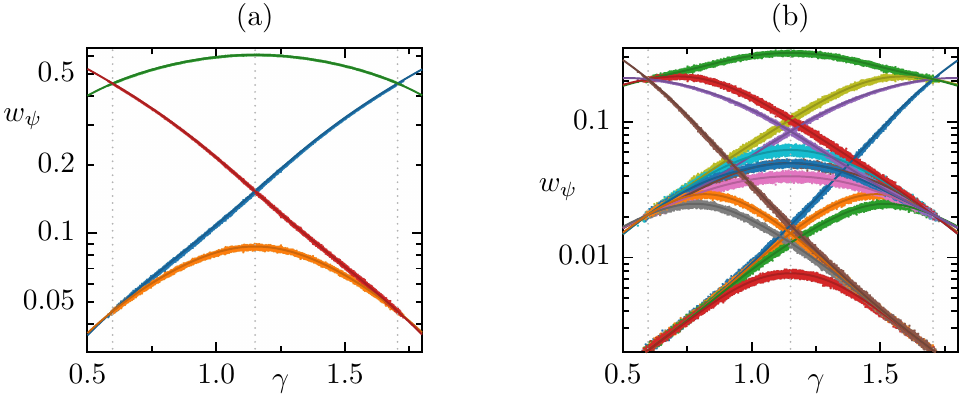}
    \caption{Comparison of $n^{2 \level}$ projected weights
        $w_{\psi_i}([\brecta_{\level,\level}])$
        of each resonance state $\psi_i$ (dots)
        and classical measures
        $\mulevel([\brecta_{\level, \level}])$ (lines)
        for $2$-baker map with $\rbold = (0.1, 1)$
        and randomization level $(\levelp, \levelq) = (\level - 1, \level)$
        for (a) $\level = 1$ and (b) $\level = 2$
        for Hilbert space dimension $N = 25600$.
        Decay rates $\gnat$, $\gtyp$, and $\ginv$ are indicated (dotted lines).
    }
    \label{fig:weights_n2}
\end{figure}

Additionally, we show the quantum-classical comparison 
for the $2$-baker map with $\rbold = (0.1, 1)$ 
for two randomization levels
in Fig.~\ref{fig:weights_n2}. 
Again, we observe excellent agreement for all decay rates.

\subsection{Jensen-Shannon divergence}

\begin{figure}[t!]
	\centering
	\includegraphics[scale=1]{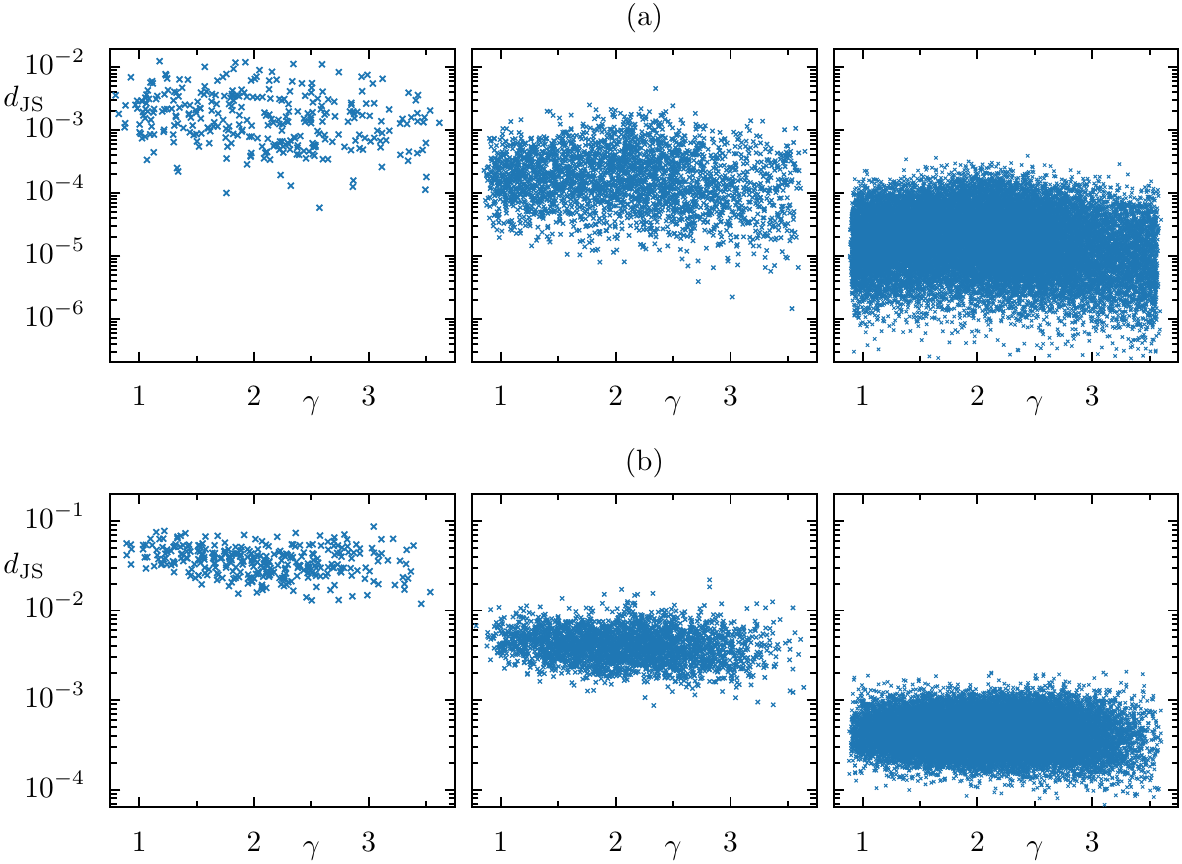}
	\caption{Jensen-Shannon divergence $\jsd$ between
		quantum and classical measures vs.\ $\gamma$
		for baker map with $\rbold = (0.2, 0.01, 1)$
		and randomization on level
		(a)
		$(\levelp, \levelq) = (0, 1)$ for increasing
		$N \in \{315, 2835, 25515\}$, and
		(b) $(\levelp, \levelq) = (1, 2)$ for
		$N \in \{324, 2835, 25515\}$
		evaluated on level $\level = \levelq$.
	}
	\label{fig:jsd}
\end{figure}

In order to further analyze the semiclassical convergence of resonance states 
we consider as in Ref.~\cite{ClaAltBaeKet2019}
the Jensen-Shannon divergence
\cite{GroBerCarRomOliSta2002}
between quantum and classical measures,
\begin{equation}
    \jsd(\muqm, \mucl) = \entropy\left(\frac{\muqm + \mucl}{2}\right)-
    \frac{\entropy(\muqm) + \entropy(\mucl)}{2},
\end{equation}
where $H$ is the Shannon entropy
\begin{equation}
	H(\mu) 
	=
	- \sum_{\brecta_{\level, \level}}
	\mu([\brecta_{\level, \level}]) \, \ln \mu([\brecta_{\level, \level}]),
\end{equation}
evaluated
on rectangles $[\brecta_{\level, \level}]$ on level $\level$.
The quantum measure $\muqm$ is determined by the projected weights
$w_{\psi_i}([\brecta_{\level,\level}])$
of a resonance state $\psi_i$ 
and the classical measure $\mucl$ is given by
$\mulevel$ for $\gamma = \gamma_i$.
The square root of $\jsd$ is a distance on the set of measures.
Weak convergence of the quantum measure to the 
classical measure implies
that for finite $\level$ the Jensen-Shannon entropy must converge to
zero in the semiclassical limit $N\rightarrow \infty$.

In Fig.~\ref{fig:jsd} we illustrate $\jsd$ versus
$\gamma$ for increasing $N$ for two different randomization levels.
For all decay rates $\gamma$
the Jensen-Shannon divergence decreases with increasing $N$.
This supports the semiclassical weak convergence conjectured
in Sec.~\ref{sec:semiclassics-conjecture}.

We remark that we tested the semiclassical
conjecture qualitatively and
quantitatively, for many different randomized baker maps,
i.e., different $n$, different $\rbold$, and also different
combinations of $\levelp$ and $\levelq$. In all cases we found
quantum-classical agreement in support of the 
semiclassical conjecture.

\section{Quantum baker map with escape}
\label{sec:usualbaker}

\begin{figure}[b!]
    \centering
    \includegraphics[scale=1]{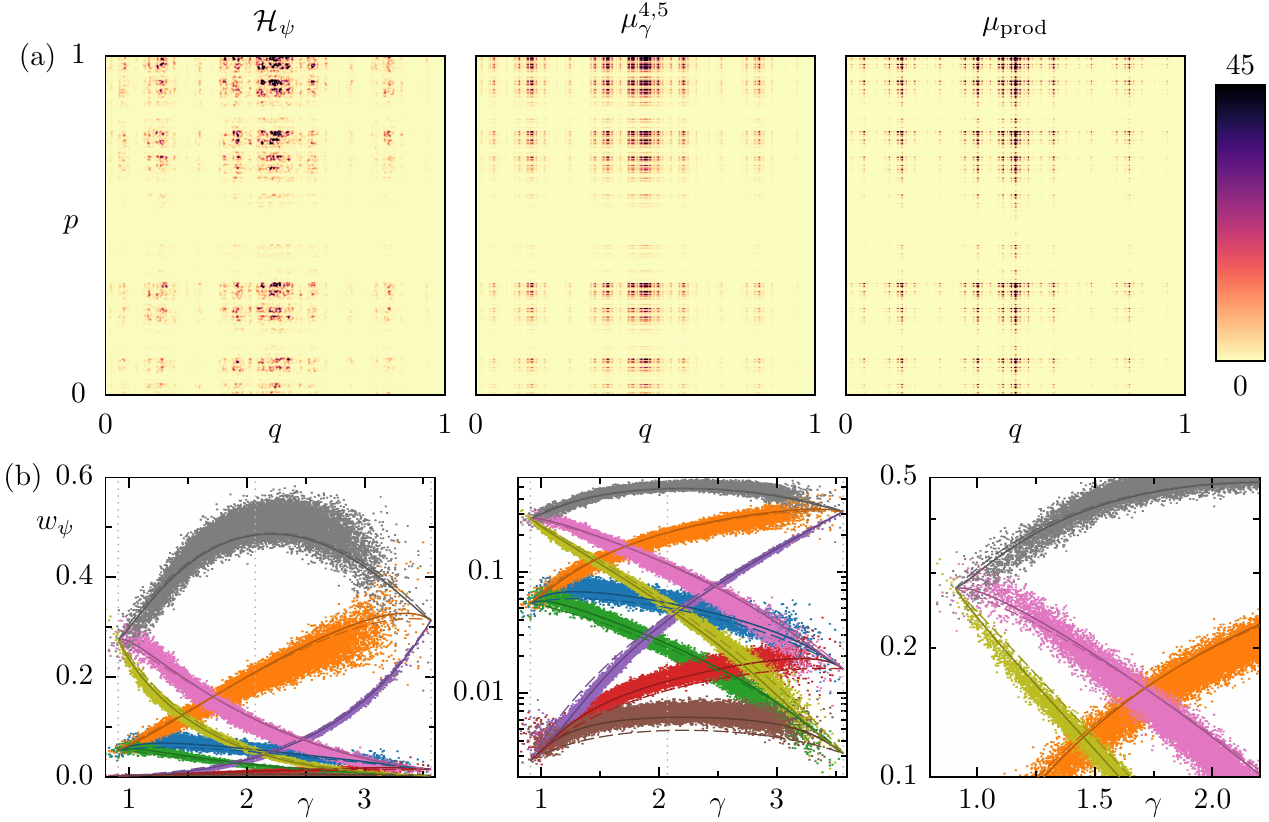}
    \caption{{Deterministic} $3$-baker map
    with $\rbold = (0.2, 0.01, 1)$ and $N = 25191$.
    (a) Husimi function
    of resonance state with decay rate closest to $\gtyp$ (left),
    corresponding
    classical measure $\mu_\gamma^{4, 5}$ (middle), 
    and product measure $\mu_\mathrm{prod}$
    based on Ref.~\cite{ClaAltBaeKet2019} (right).
    The same colormap is used for all densities.
	(b) Comparison of the 9 projected weights
	$w_{\psi_i}([\brecta_{1,1}])$
	of each resonance state $\psi_i$
	(dots)
	to classical measures
    $\mu_\gamma^{4, 5}$
    (solid lines)
	and $\mu_\mathrm{prod}$ (dashed lines).
	The same data is shown on
	linear (left) and logarithmic scale (middle), 
	as well as a magnification (right).
}
    \label{fig:baker}
\end{figure}

In this section we compare the structure of resonance states of the
deterministic quantum baker map with escape, i.e., without randomization,
to the randomized baker map with escape.
In Fig.~\ref{fig:baker}(a) in the left panel
the Husimi function of a resonance state 
of the deterministic $3$-baker map with escape $\qbmapr$, Eq.~\eqref{eq:def-qbmap},
is shown.
This is qualitatively similar to the Husimi functions 
of the randomized baker map
in Fig.~\ref{fig:husimi}(c) for small randomization regions.
Here, we compare it to the classical measures $\mulevel$ for
even smaller randomization, $(\levelp, \levelq) = (4, 5)$,
see middle panel in Fig.~\ref{fig:baker}(a).
These measures approximate the limit
$\levelp,\levelq\rightarrow\infty$, see \ref{app:convergence-measures}.
We observe qualitative agreement between the multifractal structures of both measures.

Quantitative differences can be seen 
in Fig.~\ref{fig:baker}(b),
where quantum and classical weights on the rectangles
$[\brecta_{\level,\level}]$ with $\level = 1$ are shown
for all decay rates $\gamma$.
For example, in the left panel the largest 
quantum weights do not scatter symmetrically around the corresponding
classical weight. Similar deviations can also be seen in the right panel.
Therefore the semiclassical limit measures $\mulevel$
of the randomized baker map
just approximately describe resonance states of the deterministic baker map.
Numerics suggest that this does not change,
neither for larger $N$ in the baker map with escape,
nor for classical measures with larger $\levelp, \levelq$,
see \ref{app:convergence-measures}.

Additionally, Fig.~\ref{fig:baker} shows a comparison to the 
product measures,
proposed as approximate descriptions of resonance states
in Ref.~\cite{ClaAltBaeKet2019}.
In Fig.~\ref{fig:baker}(a) in the right panel 
the product measure visually shows differences on small scales.
In Fig.~\ref{fig:baker}(b), 
where the product measure is shown as a dashed line,
it deviates substantially, e.g.,
in the middle panel for the smallest weight at intermediate $\gamma$.

There are two main conclusions from the comparison of the baker map
with and without randomization.
First of all, in both systems resonance states have qualitatively similar 
multifractal structures on scales larger than the scale
of randomization. 
In fact, the classical measures describing the resonance states of the randomized baker map 
also give the best available description for the deterministic baker map.
Secondly, however, we observe that these systems have
slightly different structures of their resonance states, 
even when the scale of randomization is chosen very small.
Thus it remains open, if and how the local random vector model
and the classical measures $\mulevel$ can
be generalized to deterministic chaotic systems.

\section{Conclusions and outlook}
\label{sec:outlook}

In conclusion, we derive a semiclassical description for resonance states
of randomized baker maps with escape based on conditional invariance and
a local random vector model.
We present numerical support which indicates that the conjectured measures
$\mulevel$ are indeed the semiclassical limit measures.

This model with randomization
acts as a bridge towards a full understanding of resonance
states in deterministic chaotic systems with escape.
The following generalizations are of interest:
(i)
The limit from partial to full escape can be
investigated for the randomized baker map
and compared with the existing approximative description
for systems with full escape~\cite{ClaKoeBaeKet2018}.
To this end the local random vector model
must be adapted to the fractal support of
resonance states in systems with full escape.
(ii)
Generalized baker maps with stripes of arbitrary width 
allow for studying non-uniformly hyperbolic systems. 
The randomization regions could either be given by finite
symbolic sequences of this map leading to different sizes
(and straightforward modifications of the local random vector model)
or be given by regions of identical sizes as in this work.  
(iii)
Randomization of other chaotic maps
and for smooth reflectivity functions, as they occur in
optical microcavities, should be studied.
Classical measures, 
based on the local random vector model and conditional invariance,
might again perfectly describe their resonance states. 
Furthermore, these classical measures might give rise to
the best available approximations for the resonance states
of deterministic maps,
similar to our observation for the baker map in
Sec.~\ref{sec:usualbaker}.

Finally, we observe that there is a quantitative difference, 
whether the multifractal structure of resonance states
ends at the scale of a Planck cell or at the scale of randomization, 
even when the resonance states are compared on much larger classical scales.
This insight should be helpful for future work on
semiclassical theories for resonance states.

\appendix
\section*{Acknowledgements}

We thank A.~B\"acker, T.~Pokart, J.R.~Schmidt,
E.-M.~Graefe, B.~Gutkin, M.~Novaes,  and D.V.~Savin
for helpful comments and discussions.
We would also like to thank A.~B\"acker, J.R.~Schmidt,
and both anonymous referees for their valuable comments and questions regarding the manuscript. 
This research is
funded by the Deutsche Forschungsgemeinschaft
(DFG, German Research Foundation) - 262765445.

\section{Substructure of randomization regions}
\label{app:subregion-structure}

\begin{figure}[t!]
	\centering
	\includegraphics[scale=1.]{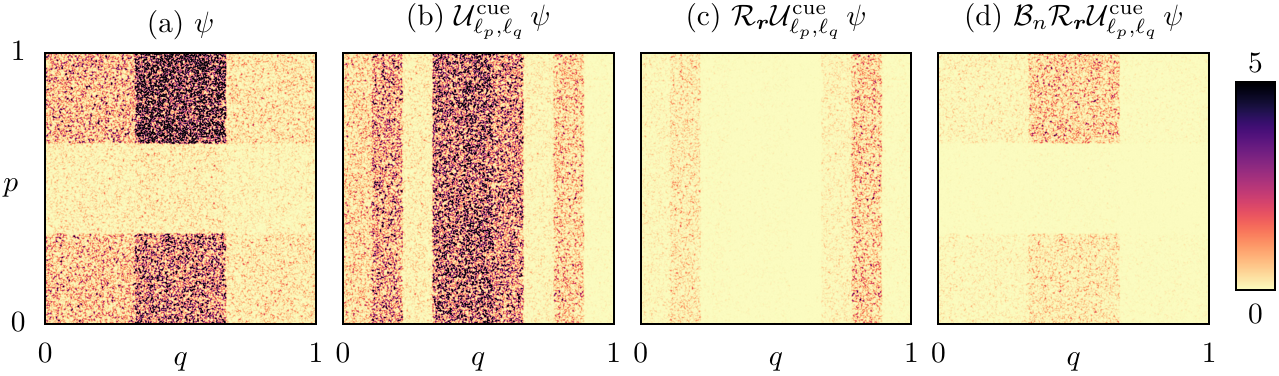}
	\caption{Husimi function of a resonance state $\psi$, as in Fig.~\ref{fig:husimi}(a, left panel)
		and sequential application of the operators
		for randomization $\urandom$, escape $\qrefl$, and baker map $\qbmap$.
	}
	\label{fig:iteration}
\end{figure}

In this section we discuss the observation that 
resonance states of the randomized baker map have
substructure within the randomization regions,
see Fig.~\ref{fig:husimi} in Sec.~\ref{sec:resonance-states}.

First we illustrate in Fig.~\ref{fig:iteration} 
how a resonance state $\psi$ evolves
under sequential application of the operators
for randomization $\urandom$, escape $\qrefl$, and baker map $\qbmap$,
that appear in the randomized baker map with esape $\qbmaprrandom$,
Eq.~\eqref{eq:def-qbmaprandom}.
The same resonance state as in Fig.~\ref{fig:husimi}(a, left panel) is used, where each of the 3 randomization regions is
divided along the $p$-direction into 3 subregions.
After application of the randomization operator, $\urandom\psi$,
each of the 3 randomization regions has again 3 subregions,
but now along the $q$-direction.
Application of the escape operator
$\qrefl$
reduces the weight in each rectangle $A_k$ by the factor $r_k$.
In particular, for $\rbold = (0.2, 0.01, 1)$ 
the weight in the middle rectangle is strongly decreased.
After application of the
quantum baker map $\qbmap$ the original 
3$\times$3-structure is recovered due to stretching in
$q$-direction and compressing in $p$-direction.
The norm, however,
is reduced by a global factor $\ue^{\gamma}$
compared to the norm of $\psi$.

Naively, one would expect to have 
no substructure in a randomization region
after application of the randomization operator $\urandom$.
This is indeed the case, if it is applied to an arbitrary state.
In Fig.~\ref{fig:iteration}, however, it is applied to an eigenstate
of the randomized baker map, \EQref{eq:def-qbmaprandom},
making a substructure in the randomization region 
before and after the application of $\urandom$ necessary.

Note that cyclic permutations of the three operators in $\qbmaprrandom$ would lead to resonance states
with phase-space structures corresponding to Fig.~\ref{fig:iteration}(b) or (c),
while the spectrum would not change.
The operators $\urandom$ and $\qrefl$ commute for $\levelq \ge 1$,
such that their exchange neither changes spectrum nor resonance states.

\section{Convergence of classical weights for $\levelp, \levelq \rightarrow \infty$}
\label{app:convergence-measures}

In the following we briefly present numerical evidence on the
weak convergence of the classical measures $\mulevel$ for increasing
$\levelp$ and $\levelq$.
Fixing a set $A\subset \torus$ 
we are interested if the limit
$\lim_{\levelp, \levelq \rightarrow\infty}
\mulevel(A)$ exists.
For simplicity $A$ is chosen as one of the rectangles $[\brecta_{\level,\level}]$ at $\level = 1$ in the following.

\begin{figure}[b!]
    \centering
    \includegraphics[scale=1]{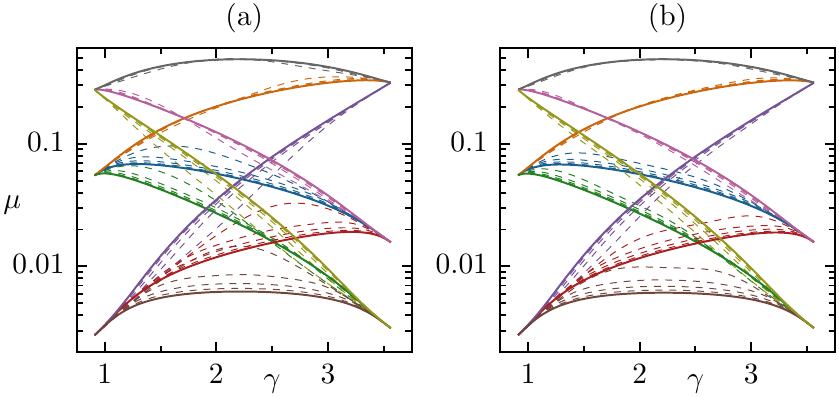}
    \caption{Convergence of classical weights
        $\mulevel([\brecta_{\level, \level}])$
        with $\level = 1$
        for $\rbold = (0.2, 0.01, 1)$ and randomization on 
        (a) $\levelp \in \{0, 1, 2, 3, 4\}$, $\levelq = \levelp + 1$,
        and (b)
        $\levelp \in \{1, 2, 3, 4, 5\}$, $\levelq = \levelp$.
        Solid lines correspond to the largest $\levelp$.
    }
    \label{fig:w-limit}
\end{figure}

In Fig.~\ref{fig:w-limit}
we illustrate the convergence of these classical weights versus $\gamma$ 
for increasing level of randomization.
We consider the cases (a) $\levelq = \levelp + 1$ and (b) $\levelq = \levelp$
for increasing $\levelp$.
We observe that these curves converge already for the small $\levelp$ presented here.
Furthermore, the classical measures for the largest $\levelp$ agree very
well between (a) and (b). This suggests, that there is a unique limit measure,
$\mu^\ast_{\gamma}$, towards which
$\mulevel$ converges weakly, regardless how
the limit $\levelp, \levelq \rightarrow \infty$ is taken. 
A proof of this observation is an open task.

\section{Random matrix model with partial escape}
\label{sec:rmt}

In this section we apply the local random vector model
combined with conditional invariance to
resonance states of a random matrix model with partial escape.
This leads to a straightforward derivation of the analytic
expressions for their structure and its dependence on $\gamma$,
given in Ref.~\cite[App.~D]{ClaKunBaeKet2021}.

\begin{figure}[b!]
    \includegraphics[scale=1]{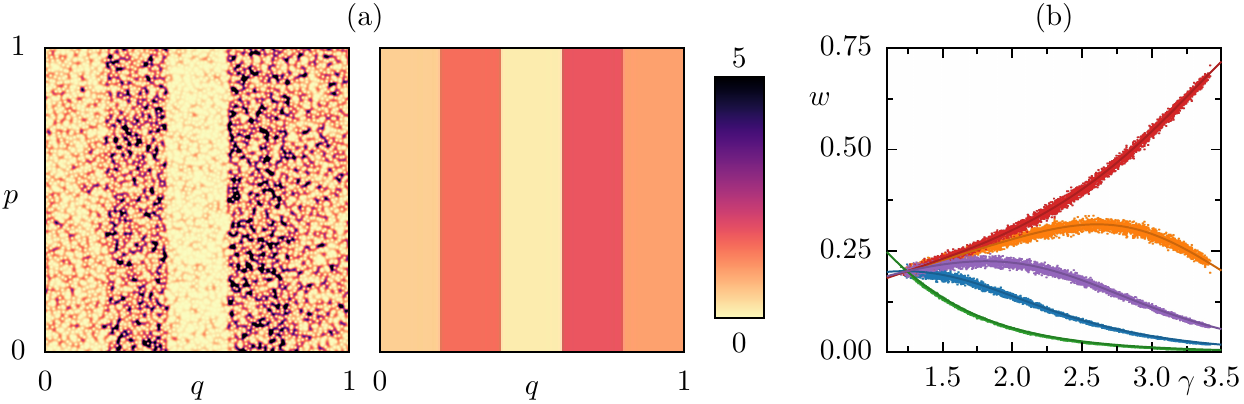}
    \caption{Random matrix with partial escape from 5 vertical stripes,
        $\rbold = (0.3, 0.03, 1, 0.01, 0.1)$ and $N = 5000$.
        (a) Husimi function of resonance state
        with decay rate closest to $\gtyp$ (left) and
        classical measure $\mu$ based on local random vector model (right),
        using the same colormap.
        (b) Comparison of the 5 projected weights
        $w_{\psi_i}([\brecta_{0,1}])$
        of each resonance state $\psi_i$
        vs. its decay rate $\gamma_i$ (dots)
        to classical weights $\mu_k$ (solid lines).
    }
    \label{fig:random}
\end{figure}

Such models are
defined 
as the composition of a random matrix and an escape operator
\cite{FyoSom1997,FyoSom2000,ZycSom2000,Bog2010,ClaKunBaeKet2021},
\begin{equation}
    \qmaprrandom
    := \urandomoo \, \qrefl
    ,
    \label{eq:def-maprandom}
\end{equation}
i.e., randomization takes place on the entire phase space, 
$\levelp = 0$, $\levelq = 0$.
Here we consider a reflectivity operator $\qrefl$
with constant escape from the rectangles $A_k$,
see Eq.~\eqref{eq:refl-operator}.
Resonance states are uniformly fluctuating
in subregions given by the rectangles $A_k$,
see Fig.~\ref{fig:random}(a), 
even though randomization takes place on the whole phase space.

We describe the structure of these resonance states 
by a classical measure $\mu$ that is constant on each rectangle $A_k$
with weight
\begin{equation}
    \mu_k
    := \mu(A_k) .
\end{equation}
According to the local random vector model, Eq.~\eqref{eq:product-lrwm},
we maximize the product of relative weights,
\begin{equation}
    \Pi(\mu)
    =
    \prod_{k=0}^{n-1}
    \frac{\mu_k}{\sum\limits_{l=0}^{n-1}\mu_l }
    =
    \prod_{k=0}^{n-1}
    \mu_k
    ,
    \label{eq:product-lrwm-rmt}
\end{equation}
where normalization $\sum_{k=0}^{n-1}\mu_k = 1$ is used
(and actually a global random vector model).
The constraint of conditional invariance follows from Eq.~\eqref{eq:def-cinv} 
applied to the entire phase space $\torus$,
\begin{equation}
    \sum\limits_{k=0}^{n-1} r_k \, \mu_k
    =
    \ue^{-\gamma} \,
    \sum\limits_{k=0}^{n-1}\mu_k
    \qquad \Leftrightarrow \qquad
    \sum\limits_{k=0}^{n-1} (\ue^{\gamma} \,r_k - 1) \, \mu_k
    =
    0
    .
    \label{eq:constraints-ci-rmt}
\end{equation}
Using the method of Lagrange multipliers we define a cost function
maximizing $\ln \Pi(\mu)$
\begin{equation}
    f(\{\mu_k\})
    =
    \sum\limits_{k=0}^{n-1} \ln \mu_k
    - \lambda 
    \left(\sum\limits_{k=0}^{n-1}\mu_k - 1
    \right)
    - \nu 
    \left(
    \sum\limits_{k=0}^{n-1} (\ue^{\gamma} \,r_k - 1) \, \mu_k
    \right)
    ,
    \label{eq:cost-rmt}
\end{equation}
with Lagrange multipliers $\lambda$ and $\nu$.
Setting the partial derivatives $(\partial/\partial \mu_k) \, f(\{\mu_k\})$ to zero
we get a set of $n$ equations,
\begin{equation}
    \frac{1}{\mu_k}
    - \lambda 
    - \nu 
    \left(\ue^{\gamma} \, r_k - 1
    \right)
    = 0
    .
    \label{eq:cond-rmt}
\end{equation}
Multiplying each equation by $\mu_k$ and summing over all $k$ gives $\lambda = n$.
The weights $\mu_k$ are thus given by,
\begin{equation}
    \mu_k = 
    \frac{1}{n + \nu (\ue^{\gamma} \, r_k - 1)}
    .
\end{equation}
This agrees with the expression for the density in Ref.~\cite[Eq.~(D10)]{ClaKunBaeKet2021},
while Eq.~\eqref{eq:constraints-ci-rmt} agrees with Ref.~\cite[Eq.~(D9)]{ClaKunBaeKet2021} and can be used for determining
the remaining Lagrange multiplier $\nu$ corresponding 
to the variable $\xi^\ast$ in Ref.~\cite{ClaKunBaeKet2021} (up to a factor $n$).

In Fig.~\ref{fig:random}
we illustrate the structure of resonance states 
of a random matrix
with escape from $n=5$ rectangles.
In (a) the Husimi function (left) and the corresponding
classical measure $\mu$ (right) are shown for $\gtyp$.
In (b)
agreement of quantum and classical weights
for all $\gamma$ and all $5$ rectangles $A_k$
is demonstrated.

\end{document}